\documentstyle[12pt,epsf]{article}

\newcommand{\1}{{\sf 1 \!\! 1}}
\newcommand{\M}{{\sf M }}
\newcommand{\da}{^\dagger}
\makeatletter
\@addtoreset{equation}{section}
\makeatother

\title{QCD as a Quantum Link Model
\footnote{This work is supported in part by funds provided by the U.S.
Department of Energy (D.O.E.) under cooperative research agreement
DE-FC02-94ER40818.}}
\author{R. Brower$^{a,b}$, S. Chandrasekharan$^a$ and U.-J. Wiese$^a$ \\ \\
$^a$ Center for Theoretical Physics, \\
Laboratory for Nuclear Science and Department of Physics \\
Massachusetts Institute of Technology (MIT) \\
Cambridge, Massachusetts 02139, U.S.A. \\ \\
$^b$ Department of Physics, Boston University \\ 
Boston, Massachusetts 02215, U.S.A. \\ \\
MIT Preprint, CTP 2623}

\begin{document} 
\maketitle
\begin{abstract} \normalsize

QCD is constructed as a lattice gauge theory in which the elements of the link
matrices are represented by non-commuting operators acting in a Hilbert space.
The resulting quantum link model for QCD is formulated with a fifth Euclidean 
dimension, whose extent resembles the inverse gauge coupling of the resulting 
four-dimensional theory after dimensional reduction. The inclusion of quarks
is natural in Shamir's variant of Kaplan's fermion method, which does not
require fine-tuning to approach the chiral limit. A rishon representation in 
terms of fermionic constituents of the gluons is derived and the quantum link 
Hamiltonian for QCD with a $U(N)$ gauge symmetry is expressed in terms of 
glueball, meson and constituent quark operators. The new formulation of QCD is 
promising both from an analytic and from a computational point of view.

\end{abstract}
 
\maketitle
 
\newpage

\section{Introduction}

Solving QCD is among the most challenging problems in theoretical physics. A
non-perturbative formulation of QCD is provided by Wilson's lattice gauge theory
\cite{Wil74}, which maps the problem to one of classical statistical mechanics.
Over the past twenty years a variety of tools have been developed to solve 
lattice field theories. At present the most powerful tool is the Monte Carlo 
simulation of the partition function of the corresponding classical statistical
mechanics system. However, the most efficient numerical algorithms for solving 
lattice QCD suffer from critical slowing down when the continuum limit is 
approached and thus exhaust even the biggest supercomputers. The use of 
improved or even ``perfect'' actions may alleviate this problem \cite{Nie96}, 
because it may allow one to extract continuum physics from rather coarse 
lattices. On the other hand, to obtain high-precision numerical data using the 
standard numerical techniques will still require large resources of computer 
power. Hence, it is certainly reasonable to look for new formulations of the 
QCD problem.

Here we propose an alternative non-perturbative approach to QCD in the framework
of quantum link models. Such models were first discussed by Horn \cite{Hor81}, 
and studied in more detail by Orland and Rohrlich \cite{Orl90}. In these models
the classical statistical mechanics problem of standard lattice gauge theory
is replaced by a problem of quantum statistical mechanics. In particular, the
classical Euclidean action is replaced by a Hamilton operator. As a 
consequence, the elements of the link matrices that are ordinary c-numbers in 
the standard formulation of lattice gauge theory, turn into non-commuting 
operators acting in a Hilbert space. Recently, quantum link models have been 
formulated with a fifth Euclidean direction and it has been shown how they are
related to ordinary 4-d Yang-Mills theories via dimensional reduction 
\cite{Cha96}. The previous work on quantum link models was limited to $U(N)$ 
and $SU(2)$ gauge groups and it was not clear how to generalize the 
construction to $SU(N)$. The quantum link models constructed in this paper have
an $SU(N)$ gauge symmetry and provide a new non-perturbative formulation of QCD.
We hope that this formulation will lead to alternative ways for attacking this 
long-standing problem, both with analytic and with numerical methods.

Quantum link QCD is quite different from the standard formulation of lattice
QCD. The main novel feature of quantum link models is that the gluonic Hilbert 
space on each link is finite. While the Hilbert space in the Hamiltonian 
formulation of standard lattice QCD contains all representations of $SU(3)$ on 
a link, the corresponding Hilbert space of an $SU(N)$ quantum link model 
consists of a single representation of $SU(2N)$. This is achieved by 
formulating the theory in $4+1$ dimensions. In order to represent full QCD, of 
course, quarks must also be included. Even in the standard formulation of 
lattice gauge theory, the Hilbert space of the quarks is finite and indeed the 
representation of quarks in quantum link QCD is not very different. However, 
the boundary conditions of the quarks in the fifth Euclidean direction must be 
carefully adjusted in order to ensure their proper dimensional reduction. 
Antiperiodic boundary conditions do not work, because they do not lead to a 
long enough fermionic correlation length in four dimensions. Periodic boundary 
conditions, on the other hand, allow an arbitrarily long 4-d fermionic 
correlation length. The resulting 4-d fermion resembles in many respects the 
standard lattice Wilson fermion. In particular, with periodic boundary 
conditions for the quarks in the fifth direction, quantum link QCD suffers from
a fine-tuning problem, when one wants to approach the chiral limit. Fortunately,
we can do better than that. 

Kaplan proposed to represent 4-d chiral fermions by working with a domain wall 
in a five-dimensional world \cite{Kap92}. In his construction the fifth 
dimension arises for completely different reasons than in quantum link models. 
However, since we already have five dimensions, it is natural to use them also
to protect the chiral symmetries of the fermions. The scheme that fits most 
naturally with quantum link models is Shamir's variant \cite{Sha93} of Kaplan's
proposal, which works with a 5-d slab (which is finite in the fifth direction) 
with open boundary conditions for the fermions at the two boundaries. Due to 
the boundary conditions, a left-handed fermion is bound to one side of the 
slab, while a right-handed fermion appears on the opposite side. Remarkably, 
the correlation length of these fermions is exponentially large in the depth of 
the slab. This is exactly what one needs for them to survive the dimensional
reduction in quantum link models. The corresponding geometry is shown in fig.1.
\begin{figure}[htb]
\epsfxsize=120mm
\epsffile{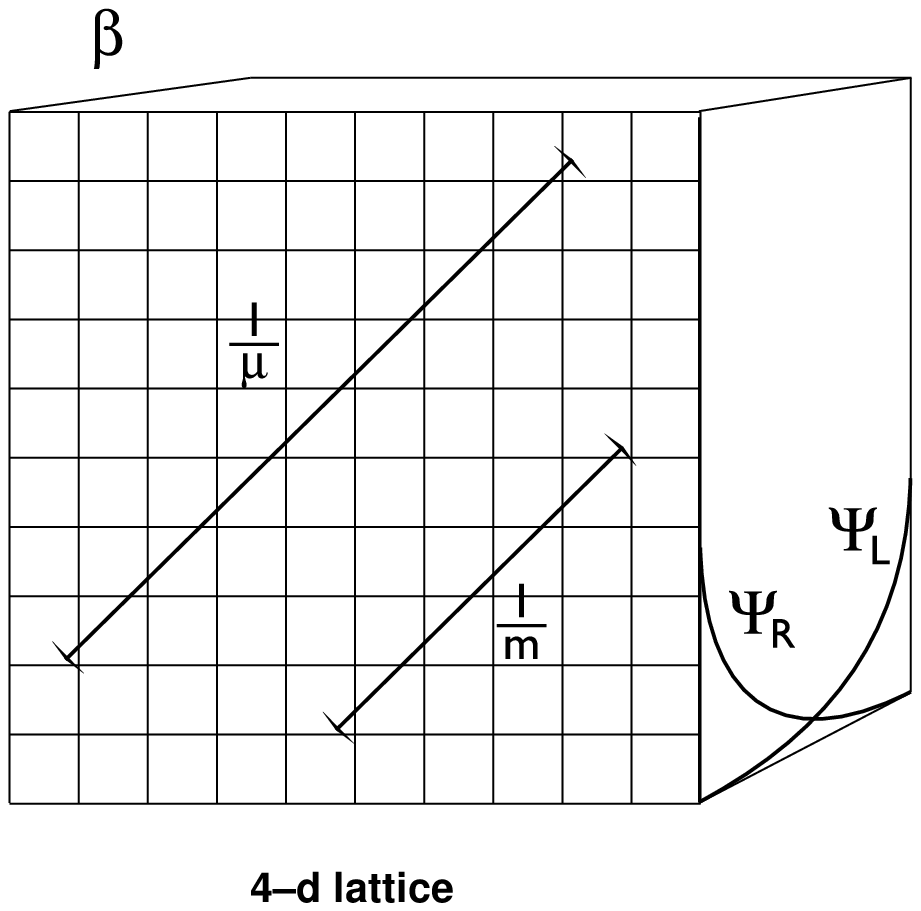}
\caption{Dimensional reduction in slab geometry: The gluonic correlation length
$1/m$ is exponential in $\beta$. Left and right-handed quarks live on the
two sides of the slab. The overlap of their wave functions $\Psi_R$ and
$\Psi_L$ induces a mass $\mu$. The corresponding correlation length $1/\mu$
is exponentially larger than $1/m$.}
\end{figure}
When Shamir's boundary conditions are used for the quarks in quantum link QCD, 
no fine-tuning is necessary to approach the chiral limit of the theory. This is
essential, because recovering the chiral properties of the quarks is one of the
main problems in numerical simulations of Wilson fermions. Recently, using the 
standard formulation of lattice gauge theory, it was demonstrated convincingly 
that Shamir's proposal is indeed practical from a computational point of view 
\cite{Blu96} and that it leads to a great improvement of the chiral properties
of quarks on the lattice. 

It is reassuring that the formulation of quantum link models in five dimensions
fits naturally with this elegant solution of the fermion doubling problem in 
vector-like theories including QCD. Should similar ideas ultimately lead to a 
lattice construction of chiral gauge theories, one may expect that the theory 
can also be formulated as a quantum link model. Once the lattice chiral fermion
problem is solved, the door may be open to a non-perturbative lattice 
formulation of supersymmetry. In that case quantum link models may offer the 
most natural framework, because they treat the Hilbert spaces of bosons and 
fermions on an equal footing. In particular, in quantum link models the bosonic
Hilbert space is also finite (on a finite lattice).

Due to the fact that the gluonic Hilbert space of quantum link QCD is finite, 
one can reformulate the theory in terms of fermionic constituents of the
gluons, which we call rishons\footnote{Rishon is Hebrew and means ``first''.} 
following jewish tradition \cite{Har79}. The rishons carry color and transform
in the fundamental representation of $SU(N)$. Each quantum link variable can be
expressed as a rishon-anti-rishon pair. The rishon dynamics is particularly 
simple. The Hamilton operator describes the hopping of rishons from one end of
a link to the other, as illustrated in fig.2. 
\begin{figure}[htb]
\epsfxsize=120mm
\epsffile{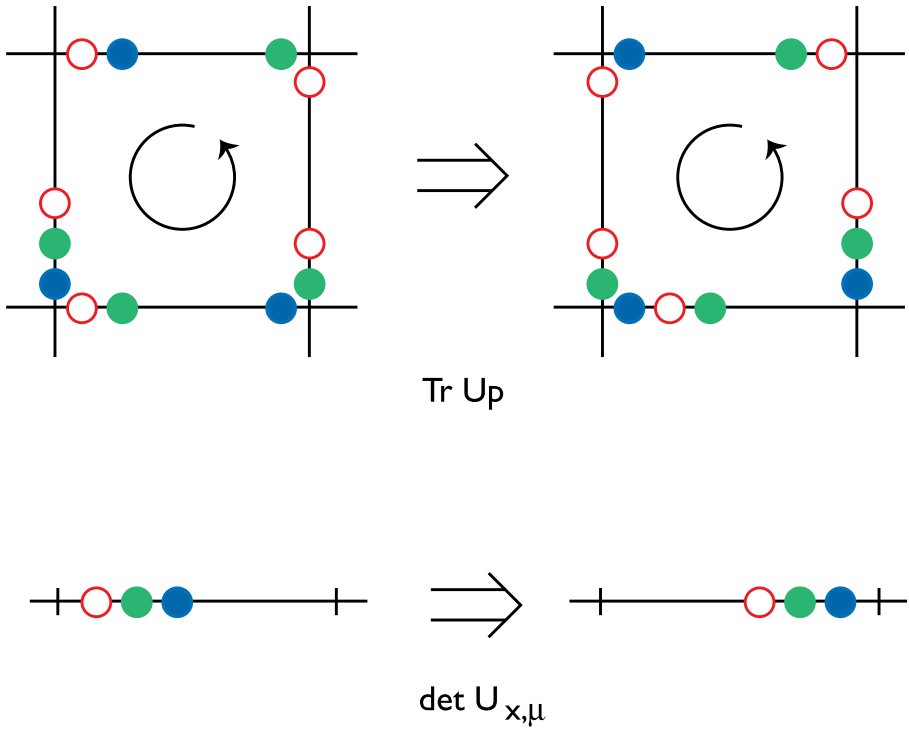}
\caption{Rishon dynamics: The trace part of the Hamiltonian induces a hopping of
rishons of various colors around a plaquette. The determinant part shifts a
color-neutral rishon baryon from one end of a link to the other.}
\end{figure}
The rishon formulation of quantum
link QCD resembles the Schwinger boson and constraint fermion constructions of 
quantum spin systems \cite{Aue94} and provides theoretical insight, that may 
lead to new analytic approaches to QCD, for example, in the large $N$ limit. 
The Hamilton operator of quantum link QCD with a $U(N)$ gauge group can also be 
expressed in terms of glueball, meson and constituent quark operators. These 
objects consist of two rishons, two quarks and a quark-rishon pair, 
respectively. It is conceivable that the theory can be bosonized in terms 
of these degrees of freedom. If large $N$ techniques can be applied 
successfully, major progress in the non-perturbative solution of QCD and other 
interesting field theories is to be expected. It is intriguing that $U(N)$
quantum link QCD has two representations --- one exclusively in terms of 
color-triplet quark and rishon fermions, the other in terms of color-singlet 
bosons.

In  contrast to standard lattice gauge theories, the configuration space of 
quantum link models is discrete and their path integral representation 
resembles that of quantum spin models. Hence, it seems plausible that highly 
efficient cluster algorithms similar to those for quantum spins \cite{Bea96} 
become available for quantum link models. For classical spin models the 
development of cluster algorithms \cite{Swe87,Wol89} gave rise to a tremendous
improvement of numerical data and has led to a satisfactory high-precision 
numerical solution of various models. Despite many attempts, it has so far not
been possible to construct efficient cluster algorithms in the standard 
formulation of lattice QCD. The 2-d classical O(3) model is in many respects 
similar to 4-d Yang-Mills theories. The most efficient way to simulate the 
$O(3)$ model is provided by the Wolff cluster algorithm \cite{Wol89}. On the 
other hand, the same model can also be obtained from the 2-d antiferromagnetic 
quantum Heisenberg model via dimensional reduction \cite{Has91}. For the
Heisenberg model a very efficient loop cluster algorithm is available
\cite{Eve93,Wie94}, which has recently been modified to work directly in the
Euclidean time continuum \cite{Bea96}. This algorithm has been used to 
simulate the correlation length of the 2-d classical O(3) model that results
from dimensional reduction \cite{Bea97a}. The quality of the numerical data
is compatible with those obtained from the Wolff algorithm applied directly to
the 2-d classical model. For example, using a finite size scaling technique,
correlation lengths of up to about 37000 lattice spacings could be handled
successfully. This suggests that, if a cluster algorithm can be constructed for
quantum link QCD, its numerical simulation should be more efficient than that
of standard lattice QCD, for which no cluster algorithm has been found.

At present, it is unclear if superior numerical methods can be developed in
the quantum link formulation of lattice QCD. However, this seems possible,
because, for example, for a $U(1)$ quantum link model a cluster algorithm
operating in the continuum of the fifth Euclidean direction has already been 
constructed \cite{Bea97}. Quantum link models with quarks resemble standard 
models of quantum statistical mechanics, as, for example, the Hubbard model. 
This may eventually lead to problems with their numerical simulation, because 
such models are known to suffer from the notorious fermion sign problem. 
However, based on a fermion cluster algorithm \cite{Wie93}, a recently proposed
method \cite{Gal96} may circumvent this difficulty. However, we postpone
further considerations of numerical simulations of quantum link QCD with 
dynamical quarks. The main issue of this paper is to present the quantum link 
formulation of QCD. 

The paper is organized as follows. In section 2 quantum link models with $U(N)$
and $SU(N)$ gauge symmetries are constructed and in section 3 it is discussed 
how these models are related to ordinary 4-d Yang-Mills theories via 
dimensional reduction. In section 4 quantum link models are reformulated in 
terms of rishons --- the fermionic constituents of the gluons. The inclusion of
quarks is discussed in section 5 within the framework of Shamir's variant of 
Kaplan's fermion method. The representation of the $U(N)$ gauge invariant 
quantum link Hamiltonian, in terms of glueball, meson and constituent quark
operators, is worked out in section 6. Finally, section 7 contains our 
conclusions.

\section{$U(N)$ and $SU(N)$ Quantum Link Models}

Quantum link models were first discussed by Horn \cite{Hor81}. He constructed 
models with $U(N)$ and $SU(2)$ gauge groups, but he pointed out that his 
construction can not be generalized to $SU(N)$. Further, the recent application 
of quantum link models to Yang-Mills theory via dimensional reduction was 
discussed only for $SU(2)$ \cite{Cha96}. Here we explicitly construct quantum 
link models with an $SU(N)$ gauge symmetry. Obviously, this is essential for 
the quantum link formulation of QCD with an $SU(3)$ gauge group.

Let us first recall the standard formulation of lattice gauge theory. In that 
case there is an $SU(N)$ matrix $u_{x,\mu}$ associated with each link $(x,\mu)$
on a 4-d hypercubic lattice. The action is given by
\begin{equation}
S[u] = - \sum_{x,\mu \neq \nu} \mbox{Tr} [u_{x,\mu} u_{x+\hat\mu,\nu} 
u_{x+\hat\nu,\mu}\da u_{x,\nu}\da],
\end{equation}
where the dagger denotes Hermitean conjugation. The action is invariant under
$SU(N)$ gauge transformations
\begin{equation}
u'_{x,\mu} = \exp(i \vec \alpha_x \cdot \vec \lambda) u_{x,\mu}
\exp(- i \vec \alpha_{x+\hat\mu} \cdot \vec \lambda),
\end{equation}
where $\vec \lambda$ is the vector of Hermitean generators of $SU(N)$ with the
usual commutation relations
\begin{equation}
[\lambda^a,\lambda^b] = 2 i f_{abc} \lambda^c, \ 
\mbox{Tr} \lambda^a \lambda^b = 2 \delta^{ab}.
\end{equation}
The path integral is given by
\begin{equation}
Z = \int {\cal D}u \exp(- \frac{1}{g^2} S[u]),
\end{equation}
where $g$ is the non-Abelian gauge coupling. Formally, the above system can be
viewed as one of classical statistical mechanics. Then the action plays the 
role of the classical Hamilton function and $g^2$ plays the role of the
temperature. The 4-d $SU(N)$ lattice gauge theory is believed to have only one
phase, in which the gluons are confined. Due to asymptotic freedom, the 
continuum limit of the lattice model corresponds to $g \rightarrow 0$.

Let us now construct a quantum link representation of $SU(N)$ lattice gauge
theory. We replace the classical action (or Hamilton function in the language
of statistical mechanics) by a quantum Hamilton operator
\begin{equation}
\label{Hamiltonian}
H = J \sum_{x,\mu \neq \nu} \mbox{Tr} [U_{x,\mu} U_{x+\hat\mu,\nu} 
U_{x+\hat\nu,\mu}\da U_{x,\nu}\da].
\end{equation}
Here the elements of the $N \times N$ link matrices $U_{x,\mu}$ are operators 
acting in a Hilbert space --- not just c-numbers like in the standard
formulation. Naturally, the dagger now represents Hermitean conjugation in both
the Hilbert space and the $N \times N$ matrix space. Note, however, that the 
trace in the above expression is only over the $N \times N$ matrix space and
not over the Hilbert space. Gauge invariance of the quantum link model requires
that the above Hamilton operator commutes with the generators $\vec G_x$ of 
infinitesimal gauge transformations at each lattice site $x$, which obey the
standard local algebra
\begin{equation}
[G^a_x,G^b_y] = 2 i \delta_{xy} f_{abc} G^c_x.
\end{equation}
Gauge covariance of a quantum link variable follows by construction if
\begin{equation}
U'_{x,\mu} = \prod_y \exp(- i \vec \alpha_y \cdot \vec G_y) U_{x,\mu} 
\prod_z \exp(i \vec \alpha_z \cdot \vec G_z) = 
\exp(i \vec \alpha_x \cdot \vec \lambda) U_{x,\mu}
\exp(- i \vec \alpha_{x+\hat\mu} \cdot \vec \lambda),
\end{equation}
where $\prod_x \exp(i \vec \alpha_x \cdot \vec G_x)$ is the unitary operator 
that represents a general gauge transformation in Hilbert space. The above 
equation implies the following commutation relations
\begin{equation}
\label{GUcommutator}
[\vec G_x,U_{y,\mu}] = \delta_{x,y+\hat\mu} U_{y,\mu} \vec \lambda -
\delta_{x,y} \vec \lambda U_{y,\mu}.
\end{equation}
In order to satisfy these relations we introduce
\begin{equation}
\vec G_x = \sum_\mu (\vec R_{x-\hat\mu,\mu} + \vec L_{x,\mu}),
\end{equation}
where $\vec R_{x,\mu}$ and $\vec L_{x,\mu}$ are generators of right and left 
gauge transformations of the link variable $U_{x,\mu}$. Suppressing the link
index $(x,\mu)$, their commutation relations take the form
\begin{equation}
[R^a,R^b] = 2 i f_{abc} R^c, \ [L^a,L^b] = 2 i f_{abc} L^c, \ [R^a,L^b] = 0,
\end{equation}
i.e. $\vec R$ and $\vec L$ generate an $SU(N)_R \otimes SU(N)_L$ algebra on 
each link. The $\vec R$ and $\vec L$ operators associated with different links 
commute with each other. The commutation relations of eq.(\ref{GUcommutator}) 
imply
\begin{equation}
[\vec R,U] = U \vec\lambda, \ [\vec L,U] = - \vec\lambda U.
\end{equation}
The above commutation relations are identical to those of the standard 
Hamiltonian formulation of lattice gauge theories. There they are the canonical
commutation relations between link variables, which play the role of
coordinates, and electric field variables $\vec R$ and $\vec L$, which play the
role of canonical conjugate momenta. In the standard formulation the 
commutation relations are realized in an infinite dimensional Hilbert space of
square integrable wave functionals of the link variables. This Hilbert space is
necessarily infinite dimensional, because one insists that the elements of a 
link matrix commute with each other. Quantum link models, on the other hand,
have a finite dimensional Hilbert space for each link, by allowing the elements
of the link fields to have non-zero commutators. It is important to realize that
this does not destroy the gauge symmetry of the problem. 

For each link the above relations can be realized by using the generators of an
$SU(2N)$ algebra, with the $SU(N)_L \otimes SU(N)_R$ algebra embedded in it. In
the fundamental representation of $SU(2N)$, for example, we can write
\begin{eqnarray}
&&\vec R = \left(\begin{array}{cc} \vec\lambda & 0 \\ 0 & 0 \end{array} 
\right), \
\vec L = \left(\begin{array}{cc} 0 & 0 \\ 0 & \vec\lambda \end{array} \right),
\nonumber \\
&&U_{ij} = \mbox{Re} U_{ij} + i \ \mbox{Im} U_{ij}, \
(U\da)_{ij} = \mbox{Re} U_{ji} - i \ \mbox{Im} U_{ji}, \nonumber \\
&&\mbox{Re} U_{ij} = \left(\begin{array}{cc} 0 & \M^{(ij)} \\ \M^{(ji)} & 0 
\end{array} \right), \
\mbox{Im} U_{ij} = \left(\begin{array}{cc} 0 & -i \M^{(ij)} \\ i \M^{(ji)} & 0 
\end{array} \right).
\end{eqnarray}
The $\M^{(ij)}$ are a set of $N^2$ matrices (one for each index pair $(ij)$) of
size $N \times N$ with $\M^{(ij)}_{kl} = \delta_{il} \delta_{jk}$. The 
commutators of real and imaginary parts of the elements of the link matrices 
take the form
\begin{eqnarray}
&&[\mbox{Re} U_{ij},\mbox{Re} U_{kl}] = [\mbox{Im} U_{ij},\mbox{Im} U_{kl}] =
- i (\delta_{ik} \ \mbox{Im} \vec \lambda_{jl} \cdot \vec R +
\delta_{jl} \ \mbox{Im} \vec \lambda_{ik} \cdot \vec L), \nonumber \\
&&[\mbox{Re} U_{ij},\mbox{Im} U_{kl}] = 
i (\delta_{ik} \ \mbox{Re} \vec \lambda_{jl} \cdot \vec R -
\delta_{jl} \ \mbox{Re} \vec \lambda_{ik} \cdot \vec L + \frac{2}{N}
\delta_{ik} \delta_{jl} T). 
\end{eqnarray}
In particular, $[U_{ij},U_{kl}] = [(U\da)_{ij},(U\da)_{kl}] = 0$. Again, we 
should emphasize that the commutation relations are local, namely all 
commutators between operators assigned to different links are zero. 

The additional generator $T$ is given by
\begin{equation}
T = \left( \begin{array}{cc} \1 & 0 \\ 0 & -\1 \end{array} \right).
\end{equation}
The real and imaginary parts of the $N^2$ matrix elements $U_{ij}$ are 
represented by $2 N^2$ Hermitean generators of $SU(2N)$. Together with the
$2 (N^2 - 1)$ generators $\vec R$ and $\vec L$ of right and left gauge
transformations and the generator $T$, these are the
\begin{equation}
2 N^2 + 2 (N^2 - 1) + 1 = 4 N^2 - 1
\end{equation}
generators of $SU(2N)$. The generator $T$ commutes with the generators of 
$SU(N)$ gauge transformations, i.e.
\begin{equation}
[T,\vec R] = [T,\vec L] = 0,
\end{equation}
but not with the elements of the link matrices, because
\begin{equation}
[T,U] = 2 U, \ [T,U\da] = - 2 U\da.
\end{equation}
This relation implies that
\begin{equation}
G_x = \frac{1}{2} \sum_\mu (T_{x-\hat\mu,\mu} - T_{x,\mu})
\end{equation}
generates an additional $U(1)$ gauge transformation, i.e.
\begin{equation}
U'_{x,\mu} = \prod_y \exp(- i \alpha_y G_y) U_{x,\mu} \prod_z 
\exp(i \alpha_z G_z) = \exp(i \alpha_x) U_{x,\mu} \exp(- i \alpha_{x+\mu}).
\end{equation}
Indeed the Hamilton operator of eq.(\ref{Hamiltonian}) is also invariant under 
the extra $U(1)$ gauge transformations and thus describes a $U(N)$ lattice 
gauge theory. The $SU(2)$ quantum link model of ref.\cite{Cha96} works with the
generators of $SO(5)$ instead of $SU(4)$. That construction does not have the 
additional $U(1)$ gauge symmetry, but it can not be generalized to $SU(N)$ as
will be explained below.

The question then arises how quantum link models can be used to represent QCD.
The answer is surprisingly simple. All one needs to do is to break the 
additional $U(1)$ gauge symmetry by adding the real part of the determinant of 
each link matrix to the Hamilton operator, such that now
\begin{equation}
\label{Hamdet}
H = J \sum_{x,\mu \neq \nu} \mbox{Tr} [U_{x,\mu} U_{x+\hat\mu,\nu} 
U_{x+\hat\nu,\mu}\da U_{x,\nu}\da]
+ J' \sum_{x,\mu} \ [\mbox{det} U_{x,\mu} + \mbox{det} U\da_{x,\mu}].
\end{equation}
Since all elements of $U_{x,\mu}$ commute with each other (although the real
and imaginary parts of an individual element do not commute) the definition of 
$\mbox{det} U_{x,\mu}$ does not suffer from operator ordering ambiguities. By
construction the above Hamiltonian is invariant under $SU(N)$ but not under the
extra $U(1)$ gauge transformations. However, there is a subtlety that needs to
be discussed. In the fundamental representation of $SU(2N)$, that was discussed
above, the operator that represents $\mbox{det} U_{x,\mu}$ turns out to be
zero. Hence, in that case even the Hamiltonian of eq.(\ref{Hamdet}) has a
$U(N)$ gauge invariance. On the other hand, since the above commutation 
relations can be realized with any representation of $SU(2N)$, we can use 
higher representations in which the determinant in general does not vanish. In 
section 4 it will become clear that the $(2N)!/(N!)^2$-dimensional 
representation of $SU(2N)$ is the smallest with a non-vanishing determinant. 
For QCD the Hilbert space of the corresponding $SU(3)$ quantum link model then 
is the direct product of 20-dimensional Hilbert spaces associated with each 
link. Specifying the state on a link thus requires just 5 bits (because 
$2^4 < 20 < 2^5$). In the standard formulation of lattice gauge theory, on the 
other hand, in the computer an $SU(3)$ link variable is usually represented by 
18 real numbers, which, for example, on a CRAY corresponds to 912 bits. 

At this point it is easy to see why there is a special construction for $SU(2)$
using a 4-dimensional representation \cite{Cha96}. Under gauge transformations
the quantum link operator $U$ transforms as an $\{N,\bar N\}$ tensor. In 
$SU(2)$ the fundamental representation $\{2\}$ and its conjugate $\{\bar 2\}$ 
are unitarily equivalent. Consequently, one can construct another quantum link
operator using $U^\dagger$ with the same $SU(2)$ gauge transformation 
properties as $U$. A linear combination of the two operators leads to a quantum
Hamiltonian with $SU(2)$ --- but not $U(2)$ --- gauge symmetry. Indeed, adding
them with equal weight gives the $SO(5)$ construction of \cite{Cha96}.

It should be clear that the Hamiltonians of eqs.(\ref{Hamiltonian}) and
(\ref{Hamdet}) represent just the simplest members of a large class of $U(N)$ 
and $SU(N)$ invariant quantum link models. Like in ordinary lattice gauge 
theories, the action (in this case the Hamiltonian) can be improved by 
including more complicated terms, for example, a six-link loop around a double
plaquette. Further, it is possible to break the extra $U(1)$ gauge symmetry by
antisymmetrized combinations of various paths connecting two lattice points ---
not just by the determinant operator on a single link. In that way one can 
construct $SU(N)$ invariant quantum link Hamiltonians even with the fundamental
representation of $SU(2N)$, but with more complicated interactions. In the 
15-dimensional representation of $SU(6)$ one can construct a more complicated 
plaquette action that leads to an $SU(3)$ invariant theory, even without the 
link determinant term. All these options may become important when the question
of the most efficient numerical treatment of quantum link models will be 
addressed.

For the moment we limit ourselves to the discussion of one natural extension of
the Hamilton operators from before. As it was discussed in ref.\cite{Cha96}, the
operators $T_{x,\mu}$, $\vec R_{x,\mu}$ and $\vec L_{x,\mu}$ resemble Abelian
and non-Abelian 5-d electric fluxes associated with the link $(x,\mu)$. On the
other hand, from a 5-d point of view the plaquette term in the above Hamilton 
operators represents the magnetic part of the energy only. Hence, it would be
natural to add a 5-d electric term as well. This corresponds to
\begin{eqnarray}
\label{Helectric}
H&=&J \sum_{x,\mu \neq \nu} \mbox{Tr} [U_{x,\mu} U_{x+\hat\mu,\nu}
U_{x+\hat\nu,\mu}\da U_{x,\nu}\da]
+ J' \sum_{x,\mu} \ [\mbox{det} U_{x,\mu} + \mbox{det} U\da_{x,\mu}]
\nonumber \\
&+&J'' \sum_{x,\mu} \ [\vec R^2_{x,\mu} + \vec L^2_{x,\mu}]
+ J''' \sum_{x,\mu} \ T^2_{x,\mu}.
\end{eqnarray}
This Hamiltonian is still gauge invariant, because $\vec R^2_{x,\mu}$,
$\vec L^2_{x,\mu}$ and $T^2_{x,\mu}$ are Casimir operators of 
$SU(N)_R \otimes SU(N)_L \otimes U(1)$, which commute with the generators of 
all gauge transformations. In the fundamental representation of $SU(2N)$ the
above combination of Casimir operators is proportional to the unit matrix and
has no effect on the dynamics. This is not the case for the higher-dimensional 
representations of $SU(2N)$. In principle, one can also include the other 
Casimir operators of $SU(N)_R \otimes SU(N)_L \otimes U(1)$ in the Hamiltonian.
The above choice of Casimir operators is most natural, because it represents 
Abelian and non-Abelian 5-d electric field energies. In the following we 
present universality arguments, which suggest that quantum link models defined 
with a 4-d Hamiltonian (describing the evolution of the system in a fifth 
Euclidean direction) get dimensionally reduced to ordinary 4-d Yang-Mills 
theories. These arguments are based on symmetry considerations and they apply 
to all $SU(N)$ invariant Hamiltonians discussed in this section. In the 
standard formulation of lattice gauge theories, one must include the electric 
terms in the action in order to get non-trivial dynamics. Since the real and 
imaginary parts of the elements of a quantum link operator do not commute, the 
dynamics of a quantum link model is already non-trivial without the electric 
terms. For simplicity, we restrict ourselves to $J'' = J''' = 0$ for the rest 
of this paper. 

\section{Reduction from Five to Four Dimensions}

As discussed in detail in ref.\cite{Cha96}, quantum link models in $4+1$
dimensions are related to ordinary 4-d gauge theories via dimensional 
reduction. The Hamilton operator of the quantum link model, which is defined
on a 4-d lattice, describes the evolution of the system in a fifth Euclidean
direction. The crucial observation is that a genuine 5-d non-Abelian gauge
theory has deconfined massless gluons and thus an infinite correlation length.
When periodic boundary conditions are imposed in the fifth direction and its 
extent $\beta$ is made finite, the extent of the extra dimension is thus
negligible compared to the correlation length. Hence, the theory appears to be 
dimensionally reduced to four dimensions. Of course, in four dimensions the 
confinement hypothesis suggests that gluons are no longer massless. Indeed, as 
it was argued in ref.\cite{Cha96}, a glueball mass
\begin{equation}
m \propto \exp(- \frac{24 \pi^2 \beta}{11 N e^2})
\end{equation}
is expected to be generated non-perturbatively. Here $e$ is the renormalized
dimensionful gauge coupling of the 5-d gauge theory. For large $\beta$ the 
gauge coupling of the dimensionally reduced 4-d theory is given by
\begin{equation}
1/g^2 = \beta/e^2.
\end{equation}
Thus the continuum limit $g \rightarrow 0$ of the 4-d theory is approached when
one sends the extent $\beta$ of the fifth direction to infinity. Hence, in 
contrast to naive expectations, dimensional reduction occurs when the extent of
the fifth direction becomes large. This is due to asymptotic freedom, which
implies that the correlation length $\xi = 1/m$ grows exponentially with 
$\beta$.

When a gauge theory is dimensionally reduced, usually the Polyakov loop in the 
extra dimension appears as an adjoint scalar field. Here we want to obtain pure
4-d Yang-Mills theory (without charged scalars) after dimensional reduction. 
This can be achieved if one does not impose Gauss' law for the states 
propagating in the fifth dimension, because the Polyakov loop is a Lagrange 
multiplier field that enforces the Gauss law. Formally, this can be realized 
simply by putting the fifth component of the gauge potential to zero, i.e.
\begin{equation}
A_5 = 0.
\end{equation}
In the infinite volume limit of the 5-d theory this restriction has no effect
on the dynamics. With finite extent in the fifth direction, however, it
deviates from the standard formulation of gauge theories. The leading terms in
the low energy effective action of the 5-d gauge theory corresponding to the
quantum link model take the form
\begin{equation}
S[A_\mu] = \int_0^\beta dx_5 \int d^4x \ \frac{1}{2 e^2}[\mbox{Tr} \ 
F_{\mu\nu} F_{\mu\nu} + \frac{1}{c^2} \mbox{Tr} \ 
\partial_5 A_\mu \partial_5 A_\mu].
\end{equation}
We expect that the QCD quantum link model leads to a 5-d gauge theory 
characterized by the ``velocity of light'' $c$. Note that here $\mu$ runs 
over 4-d indices only. At finite $\beta$ the above theory has only a 4-d gauge 
invariance, because we have fixed $A_5 = 0$, i.e. we have not imposed the 
Gauss law. On the other hand, for $\beta = \infty$ a full 5-d gauge symmetry is
recovered, although the above action then still is in $A_5 = 0$ gauge. Since we
are interested in dimensional reduction, a 4-d gauge symmetry is sufficient for
our purposes. On the level of the quantum link model, not imposing Gauss' law
is achieved by simply writing the quantum statistical partition function as
\begin{equation}
\label{gaugeZ}
Z = \mbox{Tr} \exp(- \beta H).
\end{equation}
In contrast to the standard formulation of gauge theories we have not included
a projection operator on gauge invariant states, i.e. gauge variant states also
propagate in the fifth direction. The Hamilton operator of the quantum link 
model is defined on a 4-d space-time lattice and describes the evolution of 
the system in the fifth unphysical direction. In particular, all the 
information about the physical spectrum of the 4-d theory is contained in 
correlation functions in the Euclidean time direction, which is part of the 
4-d lattice. Note that the physical Gauss law is properly imposed because the 
model does contain non-trivial Polyakov loops in the Euclidean time direction. 
In the continuum limit $g \rightarrow 0$, which we approach by increasing the 
extent $\beta$ of the fifth dimension, we are probing the low lying states in 
the spectrum of the 4-d Hamilton operator of the quantum link model. The 
space-time correlations in these unphysical states of the 4-d Hamiltonian 
contain the information about the physical spectrum. 

It is useful to think of the dimensionally reduced 4-d theory as a lattice
theory with lattice spacing $\beta c$ (which has nothing to do with the lattice
spacing of the quantum link model). In fact, one can imagine performing a block
renormalization group transformation that averages the 5-d field over cubic 
blocks of size $\beta$ in the fifth direction and of size $\beta c$ in the four
physical space-time directions. The block centers then form a 4-d space-time
lattice of spacing $\beta c$ and the effective theory of the block averaged 
5-d field is indeed a 4-d lattice theory. This is illustrated in fig.3.
\begin{figure}[htb]
\epsfxsize=120mm
\epsffile{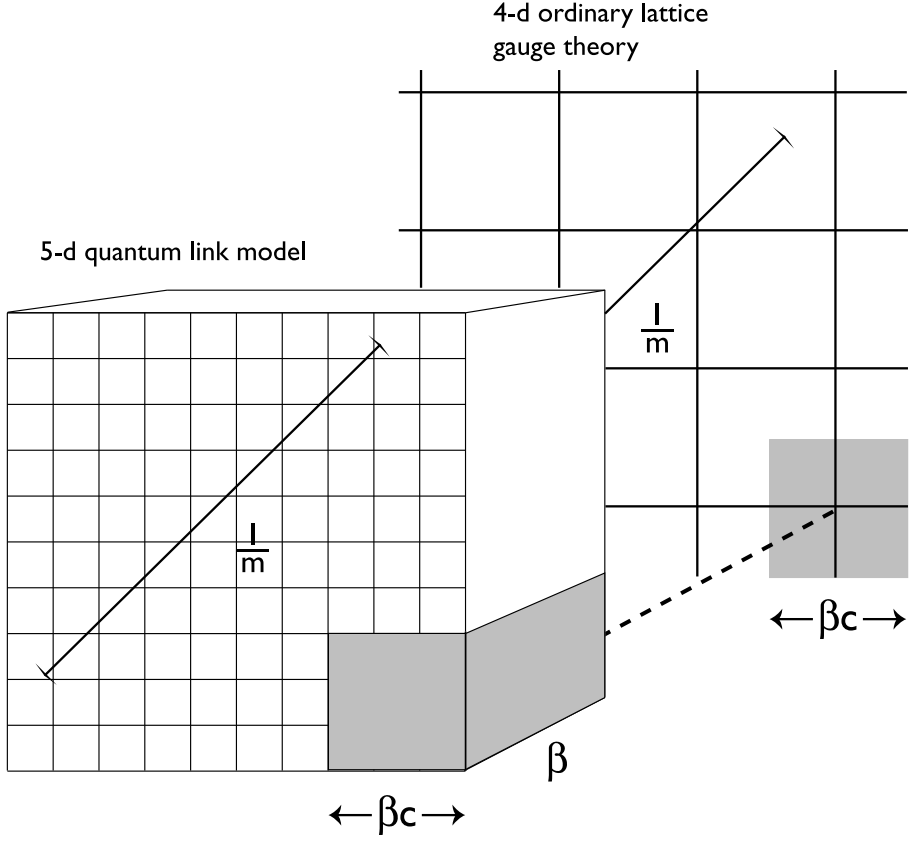}
\caption{Dimensional reduction to a 4-d lattice theory: A renormalization group
transformation that averages the 5-d field over blocks of size $\beta$ in the
fifth direction and $\beta c$ in the four space-time directions results in an
effective 4-d lattice gauge theory with lattice spacing $\beta c$.}
\end{figure}
A similar argument was first used by Hasenfratz and Niedermayer \cite{Has91} in
their study of 2-d quantum antiferromagnets.

The partition function of eq.(\ref{gaugeZ}) can be written as a 5-d path 
integral of discrete variables --- in the $SU(3)$ case the states of the 
20-dimensional representation of $SU(6)$ on each link. In many respects this 
path integral resembles that of quantum spin systems, which can be simulated by
very efficient loop cluster algorithms. Due to the discrete nature of the 
Hilbert space, one can even work directly in the continuum for the extra
Euclidean direction \cite{Bea96}. Indeed for a $U(1)$ quantum link model a
cluster algorithm operating in the continuum has already been constructed
\cite{Bea97}. It is plausible that cluster algorithms can also be constructed 
for non-Abelian quantum link models, which would allow high-precision 
simulations of gauge theories.

\section{The Rishon Representation of Quantum Link Models}

In this section we reformulate quantum link models using anticommuting 
operators describing fermionic constituents of the gluons. In contrast to 
composite models \cite{Har79}, our rishons are not physical particles, because
they are absolutely tightly bound into gluons and there is no kinetic term for
the rishons. Still, the rishons are convenient mathematical objects that 
provide theoretical insight into the physics of quantum link models. The rishon
representation is interesting both from an analytic and from a computational 
point of view. First, it may lead to a new way to attack the large $N$ limit of
QCD and other non-perturbative quantum field theories. In addition, the rishon 
formulation may also be useful in the construction of cluster algorithms.

The algebraic structure of a quantum link model is determined by the 
commutation relations derived in section 2. The Hilbert space is a direct 
product of representations of $SU(2N)$ on each link, with the generators on 
different links commuting with each other. We can thus limit ourselves to a
single link, for which the commutation relations are
\begin{eqnarray}
\label{commutators}
&&[R^a,R^b] = 2 i f_{abc} R^c, \ [L^a,L^b] = 2 i f_{abc} L^c, \ [R^a,L^b] = 
[T,R^a] = [T,L^a] = 0, \nonumber \\
&&[R^a,U_{ij}] = U_{ik} \lambda^a_{kj}, \ 
[L^a,U_{ij}] = - \lambda^a_{ik} U_{kj}, \ [T,U_{ij}] = 2 U_{ij}, \nonumber \\
&&[\mbox{Re} U_{ij},\mbox{Re} U_{kl}] = [\mbox{Im} U_{ij},\mbox{Im} U_{kl}] =
- i (\delta_{ik} \ \mbox{Im} \lambda^a_{jl} R^a +
\delta_{jl} \ \mbox{Im} \lambda^a_{ik} L^a), \nonumber \\
&&[\mbox{Re} U_{ij},\mbox{Im} U_{kl}] = 
i (\delta_{ik} \ \mbox{Re} \lambda^a_{jl} R^a -
\delta_{jl} \ \mbox{Re} \lambda^a_{ik} L^a + \frac{2}{N}
\delta_{ik} \delta_{jl} T). 
\end{eqnarray}
These relations can be realized in a representation using anticommuting rishon 
operators $c^i_{x,\pm\mu}$, $c^{i \dagger}_{x,\pm\mu}$ with color index 
$i \in \{1,2,...,N\}$. The rishon operators live at the left and right ends of 
the links and are characterized by a lattice point $x$ and a link direction 
$\pm\mu$. They obey canonical anticommutation relations
\begin{equation}
\{c^i_{x,\pm\mu},c^{j \dagger}_{y,\pm\nu}\} = \delta_{xy} 
\delta_{\pm\mu,\pm\nu} \delta_{ij}, \ 
\{c^i_{x,\pm\mu},c^j_{y,\pm\nu}\} = 0, \
\{c^{i \dagger}_{x,\pm\mu},c^{j \dagger}_{y,\pm\nu}\} = 0. 
\end{equation}
Under $SU(N)$ gauge transformations the operators $c$ and $c^\dagger$ transform
in the fundamental and anti-fundamental representation, respectively. It is 
straightforward to show that the commutation relations of 
eq.(\ref{commutators}) are satisfied when we write
\begin{eqnarray}
&&R^a_{x,\mu} = \sum_{i,j} c^{i \dagger}_{x+\hat\mu,-\mu} \lambda^a_{ij} 
c^i_{x+\hat\mu,-\mu}, \ 
L^a_{x,\mu} = \sum_{i,j} c^{i \dagger}_{x,+\mu} \lambda^a_{ij} c^i_{x,+\mu}, 
\nonumber \\
&&T_{x,\mu} = \sum_i (c^{i \dagger}_{x+\hat\mu,-\mu} c^i_{x+\hat\mu,-\mu} 
- c^{i \dagger}_{x,+\mu} c^i_{x,+\mu}), \nonumber \\
&&(U_{x,\mu})_{ij} = c^i_{x,+\mu} c^{j \dagger}_{x+\hat\mu,-\mu}, \ 
(U\da_{x,\mu})_{ij} = c^i_{x+\hat\mu,-\mu} c^{j \dagger}_{x,+\mu}.
\end{eqnarray}
Actually, the commutation relations are satisfied also when the rishons are
quantized as bosons. It is crucial to realize that all operators introduced so 
far (including the Hamiltonian) commute with the rishon number operator
\begin{equation}
{\cal N}_{x,\mu} = \sum_i (c^{i \dagger}_{x+\hat\mu,-\mu} c^i_{x+\hat\mu,-\mu}
+ c^{i \dagger}_{x,+\mu} c^i_{x,+\mu})
\end{equation}
on each individual link. Hence, we can limit ourselves to superselection 
sectors of fixed rishon number for each link. This is equivalent to working in 
a given irreducible representation of $SU(2N)$. Let us use the rishon 
representation to take a closer look at the determinant operator that we used 
to break the $U(N)$ gauge symmetry down to $SU(N)$. We have
\begin{eqnarray}
\mbox{det} U_{x,\mu}&=&\frac{1}{N!} \epsilon_{i_1 i_2 ... i_N} 
(U_{x,\mu})_{i_1 i'_1} (U_{x,\mu})_{i_2 i'_2} ... (U_{x,\mu})_{i_N i'_N} 
\epsilon_{i'_1 i'_2 ... i'_N} \nonumber \\ 
&=&\frac{1}{N!} \epsilon_{i_1 i_2 ... i_N} 
c^{i_1}_{x,+\mu} c^{i'_1 \dagger}_{x+\hat\mu,-\mu} 
c^{i_2}_{x,+\mu} c^{i'_2 \dagger}_{x+\hat\mu,-\mu} ... 
c^{i_N}_{x,+\mu} c^{i'_N \dagger}_{x+\hat\mu,-\mu} 
\epsilon_{i'_1 i'_2 ... i'_N}.
\end{eqnarray}
Due to the antisymmetry of the $\epsilon$-tensors this operator would vanish 
for bosonic rishons. For fermionic rishons, on the other hand, we have
\begin{equation}
\mbox{det} U_{x,\mu} = N! \ c^1_{x,+\mu} c^{1 \dagger}_{x+\hat\mu,-\mu} 
c^2_{x,+\mu} c^{2 \dagger}_{x+\hat\mu,-\mu} ... 
c^N_{x,+\mu} c^{N \dagger}_{x+\hat\mu,-\mu}.
\end{equation}
Only when this operator acts on a state with exactly ${\cal N} = N$ rishons
(all of a different color), it can give a non-zero contribution. In all other 
cases the determinant vanishes. This means that we can reduce the symmetry from
$U(N)$ to $SU(N)$ via the determinant only when we work with exactly 
${\cal N} = N$ fermionic rishons on each link. The number of fermion states per
link then is
\begin{equation}
\left( \begin{array}{c} 2N \\ N \end{array} \right) = \frac{(2N)!}{(N!)^2}.
\end{equation}
This is the dimension of the $SU(2N)$ representation with a totally 
antisymmetric Young tableau with $N$ boxes (arranged in a single column).

In the rishon representation the Hamilton operator takes the form
\begin{eqnarray}
\label{Hrishon}
H&=&J \sum_{x,\mu \neq \nu} \sum_{i,j,k,m} 
c^i_{x,+\mu} c^{j \dagger}_{x+\hat\mu,-\mu} 
c^j_{x+\hat\mu,+\nu} c^{k \dagger}_{x+\hat\mu+\hat\nu,-\nu} 
c^k_{x+\hat\mu+\hat\nu,-\mu} c^{m \dagger}_{x+\hat\nu,+\mu} 
c^m_{x+\hat\nu,-\nu} c^{i \dagger}_{x,+\nu} \nonumber \\
&+&J' \sum_{x,\mu} \ N! \ [c^1_{x,+\mu} c^{1 \dagger}_{x+\hat\mu,-\mu} 
c^2_{x,+\mu} c^{2 \dagger}_{x+\hat\mu,-\mu} ... 
c^N_{x,+\mu} c^{N \dagger}_{x+\hat\mu,-\mu} \nonumber \\
&+&c^1_{x+\hat\mu,-\mu} c^{1 \dagger}_{x,+\mu} 
c^2_{x+\hat\mu,-\mu} c^{2 \dagger}_{x,+\mu} ... 
c^N_{x+\hat\mu,-\mu} c^{N \dagger}_{x,+\mu}]
\nonumber \\
&=&- J \sum_{x,\mu \neq \nu} (c\da_{x+\hat\mu,-\mu} c_{x+\hat\mu,-\nu}) 
(c\da_{x+\hat\mu+\hat\nu,-\nu} c_{x+\hat\mu+\hat\nu,-\mu}) 
(c\da_{x+\hat\nu,+\mu} c_{x+\hat\nu,-\nu}) 
(c\da_{x,+\nu} c_{x,+\mu}) \nonumber \\
&+& J' \sum_{x,\mu} \ [\mbox{det} \ c_{x,+\mu} c\da_{x+\hat\mu,-\mu} 
+ \mbox{det} \ c_{x+\hat\mu,-\mu} c\da_{x,+\mu}].
\end{eqnarray}
The terms in brackets in eq.(\ref{Hrishon}) represent color neutral 
glueballs formed by two rishons located at the same corner of a 
plaquette. The plaquette part of the Hamilton operator shifts single rishons 
from one end of a link to the other and simultaneously changes their color. 
The determinant part, on the other hand, shifts an entire rishon-baryon ---
a color neutral combination of $N$ rishons --- along the link. This is
illustrated in fig.2. The rather simple rishon dynamics may facilitate new 
analytic approaches to QCD and may also be useful in numerical simulations of 
quantum link models. It is remarkable that the Hamilton operator can be 
expressed in terms of color neutral glueballs and rishon-baryons. This 
observation suggests a new approach to the large $N$ limit, which is presently 
under investigation.

\section{Quantum Link Models with Quarks}

To represent full QCD, it is essential to formulate quantum link models with 
quarks. This is more or less straightforward, although some subtleties arise
related to the dimensional reduction of fermions. Before we discuss the quantum
link formulation of full QCD, let us review the standard formulation of lattice
gauge theory with fermions. Here we concentrate on Wilson's method, although it
is straightforward to construct quantum link models with staggered fermions 
along the same lines. The standard Wilson action with quarks is given by
\begin{eqnarray}
\label{Wilsonaction}
S[\bar\psi,\psi,u]
&=&- \sum_{x,\mu \neq \nu} \mbox{Tr} [u_{x,\mu} u_{x+\hat\mu,\nu} 
u\da_{x+\hat\nu,\mu} u\da_{x,\nu}] \nonumber \\
&+&\frac{1}{2} \sum_{x,\mu} \ [\bar\psi_x \gamma_\mu u_{x,\mu} \psi_{x+\hat\mu}
- \bar\psi_{x+\hat\mu} \gamma_\mu u\da_{x,\mu} \psi_x] 
+ M \sum_x \bar\psi_x \psi_x \nonumber \\
&+&\frac{r}{2} \sum_{x,\mu} \ [2 \bar\psi_x \psi_x 
- \bar\psi_x u_{x,\mu} \psi_{x+\hat\mu}
- \bar\psi_{x+\hat\mu} u\da_{x,\mu} \psi_x].
\end{eqnarray}
Here $\bar\psi_x$ and $\psi_x$ are independent Grassmann valued spinors
associated with the lattice site $x$, $\gamma_\mu$ are Dirac matrices and
$M$ is the bare quark mass. The term proportional to $r$ is the Wilson term 
that removes unwanted lattice fermion doublers at the expense of explicitly 
breaking chiral symmetry. In order to reach the continuum limit the bare mass 
$M$ must be tuned appropriately.

The guiding principle in the formulation of quantum link models is to replace 
the classical action of the standard formulation by a Hamilton operator that
describes the evolution of the system in a fifth Euclidean direction. For the
quarks we must also replace $\bar\psi_x$ by $\Psi\da_x \gamma_5$. Hence, the 
full QCD quantum link Hamiltonian is given by
\begin{eqnarray}
\label{QCDaction}
H&=&J \sum_{x,\mu \neq \nu} \mbox{Tr} [U_{x,\mu} U_{x+\hat\mu,\nu} 
U\da_{x+\hat\nu,\mu} U\da_{x,\nu}]
+ J' \sum_{x,\mu} \ [\mbox{det} U_{x,\mu} + \mbox{det} U\da_{x,\mu}] 
\nonumber \\
&+&\frac{1}{2} \sum_{x,\mu} \ [\Psi\da_x \gamma_5 \gamma_\mu U_{x,\mu} 
\Psi_{x+\hat\mu}
- \Psi\da_{x+\hat\mu} \gamma_5 \gamma_\mu U\da_{x,\mu} \Psi_x] 
+ M \sum_x \Psi\da_x \gamma_5 \Psi_x \nonumber \\
&+&\frac{r}{2} \sum_{x,\mu} \ [2 \Psi\da_x \gamma_5 \Psi_x 
- \Psi\da_x \gamma_5 U_{x,\mu} \Psi_{x+\hat\mu}
- \Psi\da_{x+\hat\mu} \gamma_5 U\da_{x,\mu} \Psi_x].
\end{eqnarray}
Here $\Psi\da_x$ and $\Psi_x$ are quark creation and annihilation operators
with canonical anticommutation relations
\begin{equation}
\{\Psi^{i a \alpha}_x,\Psi^{j b \beta \dagger}_y\} = 
\delta_{xy} \delta_{ij} \delta_{ab} \delta_{\alpha \beta}, \
\{\Psi^{i a \alpha}_x,\Psi^{j b \beta}_y\} = 
\{\Psi^{i a \alpha \dagger}_x,\Psi^{j b \beta \dagger}_y\} = 0,
\end{equation}
where $(i,j)$, $(a,b)$ and $(\alpha,\beta)$, which have been suppressed in 
eq.(\ref{QCDaction}), are color, flavor and Dirac indices, respectively. Of 
course, we have again replaced the classical link variables $u_{x,\mu}$ by 
quantum link operators $U_{x,\mu}$. The generator of an $SU(N)$ gauge
transformation now takes the form
\begin{equation}
\vec G_x = \sum_\mu (\vec R_{x-\hat\mu,\mu} + \vec L_{x,\mu})
+ \Psi\da_x \vec \lambda \Psi_x,
\end{equation}
and it is again straightforward to show that $H$ commutes with $\vec G_x$ for
all $x$.

The same universality arguments that were used before now suggest that the 
effective action of the corresponding 5-d gauge theory (with $A_5 = 0$) 
takes the form
\begin{eqnarray}
\label{effaction}
S[\bar\psi,\psi,A_\mu]&=&\int_0^\beta dx_5 \int d^4x \ 
\{ \frac{1}{2 e^2} [\mbox{Tr} F_{\mu\nu} F_{\mu\nu} 
+ \frac{1}{c^2} \mbox{Tr} \partial_5 A_\mu \partial_5 A_\mu] \nonumber \\
&+&\bar\psi[\gamma_\mu (A_\mu + \partial_\mu) + m_q + 
\frac{1}{c'} \gamma_5 \partial_5] \psi \}.
\end{eqnarray}
The ``velocity of light'' $c'$, which characterizes the propagation of a 
quark in the fifth direction, is in general different from the corresponding
quantity $c$ for the gluons, because the quantum link formulation has no 
symmetry between the four physical space-time directions and the extra fifth 
direction. This is no problem, because we are only interested in the 4-d 
physics after dimensional reduction.

To ensure the proper dimensional reduction of the quarks, their boundary
conditions in the fifth direction must be chosen appropriately. The standard
antiperiodic boundary conditions, which are dictated by thermodynamics in the
Euclidean time direction, would lead to Matsubara modes, 
$p_5 = 2 \pi (n_5 + \frac{1}{2})/\beta$, which would limit the physical
correlation length of the dimensionally reduced fermion to $O(\beta c')$. The
confinement physics of the induced 4-d gluon theory, on the other hand, takes
place at a correlation length which is growing exponentially with $\beta$. In 
fact, $\beta$ plays the role of the lattice spacing of the dimensionally 
reduced theory. Quarks with antiperiodic boundary conditions in the fifth 
direction would hence remain at the cut-off and the dimensionally reduced 
theory would still be a Yang-Mills theory without quarks. Once this problem is 
understood, one possible solution is obvious. One can simply choose periodic 
boundary conditions for the quarks in the fifth direction. This gives rise to
a Matsubara mode, $p_5 = 0$, that survives dimensional reduction. Since the 
extent of the fifth direction has nothing to do with the inverse temperature 
(which is the extent of the Euclidean time direction), one could indeed choose 
the boundary condition in this way. 

Fortunately, we can do better. The above scenario with periodic boundary
conditions for the quarks would suffer from the same fine-tuning problem as the
original Wilson fermion method. The bare quark mass would have to be adjusted 
very carefully in order to reach the chiral limit. In practice this is a great 
problem in numerical simulations. This problem has been solved very elegantly 
in Shamir's variant \cite{Sha93} of Kaplan's fermion proposal \cite{Kap92}. 
Kaplan studied the physics of a 5-d system of fermions, which is always 
vector-like, coupled to a 4-d domain wall that manifests itself as a 
topological defect. The key observation is that under these conditions a zero 
mode of the 5-d Dirac operator appears as a bound state localized on the domain
wall. From the point of view of the 4-d domain wall, the zero mode represents a
massless chiral fermion. The original idea was to construct lattice chiral 
gauge theories in this way. Shamir has pointed out that the same mechanism can 
solve the lattice fine-tuning problem of the bare fermion mass in vector-like 
theories including QCD. He also suggested a variant of Kaplan's method that has 
several technical advantages and that turns out to fit very naturally with the
construction of quantum link QCD. In quantum link models we already have a 
fifth direction for reasons totally unrelated to the chiral symmetry of 
fermions. We will now follow Shamir's proposal and use the fifth direction to
solve the fine-tuning problem that we would have with periodic boundary 
conditions for the quarks. 

The essential technical simplification compared to Kaplan's original proposal 
is that one now works with a 5-d slab of finite size $\beta$ with open boundary
conditions for the fermions at the two sides. This geometry limits one to 
vector-like theories, because now there are two zero modes --- one at each 
boundary --- which correspond to one left and one right-handed fermion in four
dimensions. This set-up fits naturally with our construction of quantum link 
QCD. In particular, the evolution of the system in the fifth direction is still
governed by the Hamilton operator of eq.(\ref{QCDaction}). The only (but 
important) difference to Wilson's fermion method is that now $r < 0$. Of 
course, one could also obtain a left and a right-handed fermion by using a 
domain wall and an anti-wall with otherwise periodic boundary conditions. In 
that case the Hamiltonian of eq.(\ref{QCDaction}) would have to be modified in 
an $x_5$-dependent way. Shamir's method is more economical and concentrates on 
the essential topological aspects, which are encoded in the boundary conditions
for the fermions in the fifth direction. It is important that in Shamir's 
construction one also puts $A_5 = 0$. There are some minor differences between 
the implementations of the method in the standard formulation of lattice gauge
theory and in quantum link models. In the standard formulation one works with
a 4-d gauge field, which is constant in the fifth direction. In quantum link
QCD this is not possible, because the non-trivial dynamics in the fifth 
direction turns the discrete states of quantum links into the continuous
degrees of freedom of physical gluons. However, it is still true that the
physical gluon field is essentially constant in the fifth direction, because
its correlation length grows exponentially with $\beta$. This is important for
the generation of the fermionic zero modes at the two sides of the 5-d slab.

The partition function of the theory with open boundary conditions for the
quarks and with periodic boundary conditions for the gluons is simply given by
\begin{equation}
Z = \mbox{Tr} \langle 0|\exp(- \beta H)|0\rangle.
\end{equation}
Here the trace extends only over the gluonic Hilbert space of the quantum link
model, thus implementing periodic boundary conditions for the gluons. We
decompose the quark spinor into left and right-handed components
\begin{equation}
\Psi_x = \left( \begin{array}{c} \Psi_{Rx} \\ \Psi\da_{Lx} \end{array} \right).
\end{equation}
Open boundary conditions for the fermions are realized by taking the expectation
value of $\exp(- \beta H)$ in the Fock state $|0\rangle$, which is annihilated
by all right-handed $\Psi_{R x}$ and by all left-handed $\Psi_{L x}$ 
\cite{Fur95}. As a result, there are no left-handed quarks at the boundary at 
$x_5 = 0$ and there are no right-handed quarks at the boundary at 
$x_5 = \beta$. Of course, unlike periodic or antiperiodic boundary conditions, 
open boundary conditions for the fermions break translation invariance in the 
fifth direction. Through the interaction between quarks and gluons, this 
breaking also affects the gluonic sector. This is no problem, because we are 
only interested in the 4-d physics after dimensional reduction. For that it is 
essential that both quarks and gluons have zero modes, which is indeed the case
with the boundary conditions from above. 

Of course, as we have argued before, the gluonic correlation length is not
truly infinite as long as $\beta$ is finite, but --- due to confinement --- it 
is exponentially large. As we will see now, the same is true for the quarks, 
but for a totally different reason. In fact, already free quarks pick up an 
exponentially small mass due to tunneling between the two boundaries, which 
mixes left and right-handed states, and thus breaks chiral symmetry explicitly.
To understand this, let us consider the actual Hamiltonian ${\cal H}(\vec p)$, 
which describes the evolution of free quarks with spatial lattice momentum 
$\vec p$ in time (rather than $H$, which describes the evolution of the system 
in the fifth Euclidean direction)
\begin{equation}
{\cal H}(\vec p) = \gamma_4 [\gamma_5 \partial_5 + M + 
\frac{r}{2} \sum_i (2 \sin\frac{p_i}{2})^2 - i \sum_i \gamma_i \sin p_i].
\end{equation}
The wave function of a stationary quark state with energy $E(\vec p)$ is 
determined by
\begin{equation}
\label{Dirac}
{\cal H}(\vec p) \Psi(\vec p,x_5) = E(\vec p) \Psi(\vec p,x_5).
\end{equation}
Let us consider states with lattice momenta $p_i = 0,\pi$. The state with
$\vec p = \vec 0$ describes a physical quark at rest, while the other states
correspond to doubler fermions. Denoting the number of non-zero momentum 
components $p_i = \pi$ by
\begin{equation}
n = \sum_i \sin^2\frac{p_i}{2},
\end{equation}
the above Dirac equation (\ref{Dirac}) takes the form
\begin{eqnarray}
&&- \partial_5 \Psi_L(\vec p,x_5) + (M + 2 n r) \Psi_L(\vec p,x_5) = E(\vec p)
\Psi_R(\vec p,x_5), \nonumber \\
&&\partial_5 \Psi_R(\vec p,x_5) + (M + 2 n r) \Psi_R(\vec p,x_5) = E(\vec p) 
\Psi_L(\vec p,x_5).
\end{eqnarray}
Due to the boundary conditions introduced before, we must solve these equations 
with $\Psi_L(\vec p,0) = \Psi_R(\vec p,\beta) = 0$. Inserting the ansatz
\begin{equation}
\Psi_L(\vec p,x_5) = A \sinh \alpha x_5, \ \Psi_R(\vec p,x_5) =
\pm A \sinh \alpha (x_5 - \beta),
\end{equation}
immediately gives
\begin{equation}
\alpha = (M + 2 n r) \tanh \alpha \beta, \ 
M + 2 n r = \pm E(\vec p) \cosh \alpha \beta.
\end{equation}
This equation has a normalizable solution only if $M + 2 n r > 0$. In that 
case, for large $\beta$ this implies $\alpha = \pm (M + 2 n r)$ and hence
\begin{equation}
E(\vec p) = \pm 2 (M + 2 n r) \exp(- (M + 2 n r) \beta).
\end{equation}
The important observation is that the bare mass parameter $M$ is not the 
physical quark mass, although we have considered an unrenormalized free theory.
Let us first consider the doubler fermions, which are characterized by $n > 0$.
It is then possible to choose $r$ such that $M + 2 n r < 0$, for which no
normalizable solution exists. Thus, for $r < - M/2$, i.e. for a sufficiently 
strong Wilson-term with an unconventional sign, the doubler fermions are 
removed from the physical spectrum. The mass of the physical fermion 
(characterized by $n = 0$) is
\begin{equation}
\mu = |E(\vec 0)| = 2 M \exp(- M \beta),
\end{equation}
which is exponentially small in $\beta$. This situation is illustrated in fig.2.
The above result suggests how the fine-tuning problem of the fermion mass can 
be avoided. The confinement physics of quantum link QCD in the chiral limit 
takes place at a length scale
\begin{equation}
\frac{1}{m} \propto \exp(\frac{24 \pi^2 \beta}{(11 N - 2 N_f) e^2}),
\end{equation} 
which is determined by the 1-loop coefficient of the $\beta$-function of QCD 
with $N_f$ massless quarks and by the 5-d gauge coupling $e$. As long as one
chooses
\begin{equation}
M > \frac{24 \pi^2}{(11 N - 2 N_f) e^2},
\end{equation}
the chiral limit is reached automatically when one approaches the continuum
limit by making $\beta$ large. For a given value of $r$ one is limited by
$M < - 2 r$ (note that $r < 0$). On the other hand, one can always choose $J$
(and thus $e^2$) such that the above inequality is satisfied.

Of course, we also want to be able to work at non-zero quark masses. Following 
Shamir, we do this by modifying the boundary conditions for the quarks in the
fifth direction. Instead of using $\psi_L(\vec x,x_4,0) = 
\psi_R(\vec x,x_4,\beta) = 0$ we now demand
\begin{equation}
\label{boundary}
\psi_L(\vec x,x_4,0) = - \frac{m_q}{2 M} \psi_L(\vec x,x_4,\beta), \
\psi_R(\vec x,x_4,\beta) = - \frac{m_q}{2 M} \psi_R(\vec x,x_4,0),
\end{equation}
where $m_q$ is a mass parameter. This reduces to the previous boundary 
condition for $m_q = 0$, while it corresponds to antiperiodic boundary 
conditions for $m_q = 2 M$, and to periodic boundary conditions for 
$m_q = - 2 M$. Solving the above equations with the new boundary condition 
indeed yields physical quarks of mass $m_q$ in the continuum limit 
$\beta \rightarrow \infty$ as long as $m_q \ll M$. It has been shown in 
ref.\cite{Fur95} that in the interacting theory $m_q$ is only multiplicatively 
renormalized. On the level of the partition function 
\begin{equation}
\label{partitionfunction}
Z = \mbox{Tr}[\exp(- \beta H) {\cal O}(m_q)]
\end{equation}
the new boundary condition manifests itself by a mass-dependent operator
\begin{equation}
{\cal O}(m_q) = \prod_x 
(\Psi_{Rx} \Psi\da_{Rx} + \frac{m_q}{2 M} \Psi\da_{Rx} \Psi_{Rx})
(\Psi_{Lx} \Psi\da_{Lx} + \frac{m_q}{2 M} \Psi\da_{Lx} \Psi_{Lx}),
\end{equation}
which was constructed in ref.\cite{Fur95}. Note that in 
eq.(\ref{partitionfunction}) the trace is both over the gluonic and over the 
fermionic Hilbert space. In the chiral limit $m_q = 0$ the operator 
${\cal O}(m_q)$ reduces to a projection operator $|0\rangle \langle 0|$ on the 
Fock state introduced before. For $m_q = 2M$, i.e. for antiperiodic boundary 
conditions, the operator ${\cal O}(m_q)$ becomes the unit operator and the 
partition function reduces to the expression well-known from thermodynamics.

Like any massive fermion action in odd dimensions, the effective action of 
eq.(\ref{effaction}) is not parity symmetric. Hence, there is the potential 
danger that the corresponding symmetry breaking terms make their way into the 
dimensionally reduced 4-d theory. Indeed the topological $\theta$-vacuum term 
violates parity in four dimensions. For the moment we are interested in QCD 
with $\theta = 0$ and therefore in a parity symmetric theory. Later, we will 
ask how to incorporate $\theta$. We will now show that the dimensionally 
reduced theory is indeed parity invariant. For simplicity this discussion will 
be based on the low energy continuum effective action of eq.(\ref{effaction}). 
It is, however, straightforward to apply the same arguments to the lattice 
action in the path integral that one obtains from the Hamiltonian of 
eq.(\ref{QCDaction}). The 5-d effective action is not invariant under parity as
it is usually defined in five dimensions. Since we are interested in 4-d 
physics, this definition of parity is, however, not relevant for us. Instead, 
we now define another transformation that leaves the 5-d action invariant and 
reduces to ordinary parity in the dimensionally reduced 4-d theory. Let us 
consider
\begin{eqnarray}
&&\psi'(\vec x,x_4,x_5) = \gamma_4 \psi(- \vec x,x_4,\beta - x_5), \
\bar\psi'(\vec x,x_4,x_5) =  \bar\psi(- \vec x,x_4,\beta - x_5) 
\gamma_4, \nonumber \\
&&A'_i(\vec x,x_4,x_5) = - A_i(- \vec x,x_4,\beta - x_5), \
A'_4(\vec x,x_4,x_5) = A_4(- \vec x,x_4,\beta - x_5). \nonumber \\ \
\end{eqnarray}
It is straightforward to show that this is indeed a symmetry of the 5-d 
effective action. Note that we have changed $x_5$ to $\beta - x_5$. This change
is invisible from the point of view of the dimensionally reduced 4-d theory. 
In fact, in the 4-d world the above transformation reduces to ordinary parity.
However, exchanging $x_5$ and $\beta - x_5$ is essential as far as the 5-d 
theory is concerned. This operation exchanges the two domain walls and hence
the left and right-handed fermions that are bound to them. This is indeed what
parity is supposed to do.

A non-perturbative formulation of QCD would be incomplete without a discussion 
of the vacuum angle $\theta$. Although in the real world this parameter is 
indistinguishable from zero, it represents a possible parity breaking effect in
the QCD Lagrangian. In the standard formulation of lattice QCD such a term 
could be added via the topological charge, for which a lattice regularized 
expression with adequate topological properties exists \cite{Lue83}. In quantum
link QCD this would not be possible, because the discrete states of the quantum
link operators can not directly encode topological properties. Here we propose 
to include $\exp(i \theta)$ as the determinant of the fermion mass matrix and
hence via the boundary condition of eq.(\ref{boundary}). When one uses a 
fermion formulation which does not require fine-tuning in the chiral limit, 
this is the most natural thing to do, even in the standard formulation of 
lattice QCD.

\section{Glueballs, Mesons and Constituent Quarks}

In this section we express the Hamiltonian of the $U(N)$ quantum link model 
with quarks in terms of color neutral operators. Gauge invariance requires that
these operators are local bilinear combinations of rishons and quarks, which we
refer to as glueballs, mesons and constituent quarks. All these objects ---
including the constituent quarks --- are bosons. In fact, the resulting 
representation of the theory can be viewed as a first step towards a 
bosonization of $U(N)$ QCD. The determinant term in the $SU(N)$ quantum link 
Hamiltonian would give rise to an additional color-singlet rishon-baryon 
consisting of $N$ rishons. For odd $N$ this object would hence be a fermion. To
avoid complications related to these objects we limit ourselves to $U(N)$ in 
this section.
 
As in the Yang-Mills case, we represent the quantum link Hamiltonian of $U(N)$ 
QCD in terms of rishons
\begin{eqnarray}
H&=&- J \sum_{x,\mu \neq \nu} (c\da_{x+\hat\mu,-\mu} c_{x+\hat\mu,+\nu})
(c\da_{x+\hat\mu+\hat\nu,-\nu} c_{x+\hat\mu+\hat\nu,-\mu}) 
(c\da_{x+\hat\nu,+\mu} c_{x+\hat\nu,-\nu}) 
(c\da_{x,+\nu} c_{x,+\mu}) \nonumber \\
&+&\frac{1}{2} \sum_{x,\mu} \ [(\Psi\da_x c_{x,+\mu}) 
\gamma_5 \gamma_\mu (c\da_{x+\hat\mu,-\mu} \Psi_{x+\hat\mu})
- (\Psi\da_{x+\hat\mu} c_{x+\hat\mu,-\mu}) \gamma_5 \gamma_\mu 
(c\da_{x,+\mu} \Psi_x)] \nonumber \\ 
&+&M \sum_x \Psi\da_x \gamma_5 \Psi_x \nonumber \\
&+&\frac{r}{2} \sum_{x,\mu} \ [2 \Psi\da_x \gamma_5 \Psi_x 
- (\Psi\da_x c_{x,+\mu}) \gamma_5 (c\da_{x+\hat\mu,-\mu} \Psi_{x+\hat\mu})
- (\Psi\da_{x+\hat\mu} c_{x+\hat\mu,-\mu}) \gamma_5 (c\da_{x,+\mu} \Psi_x)].
\nonumber \\ \
\end{eqnarray}
Note that some of the terms in brackets now represent color neutral bosonic
constituent quarks formed by a quark and a rishon located at the same end 
of a link. In fact, the Hamiltonian can be expressed as
\begin{eqnarray}
H&=&- J \sum_{x,\mu \neq \nu} \Phi_{x+\hat\mu,-\mu,+\nu}
\Phi_{x+\hat\mu+\hat\nu,-\nu,-\mu} \Phi_{x+\hat\nu,+\mu,-\nu}
\Phi_{x,+\nu,+\mu} \nonumber \\
&+&\frac{1}{2} \sum_{x,\mu} \ [Q\da_{x,+\mu} \gamma_5 \gamma_\mu 
Q_{x+\hat\mu,-\mu}
- Q\da_{x+\hat\mu,-\mu} \gamma_5 \gamma_\mu Q_{x,+\mu}] 
+ M \sum_x \mbox{Tr}(M_x \gamma_5) \nonumber \\
&+&\frac{r}{2} \sum_{x,\mu} \ [2 \mbox{Tr}(M_x \gamma_5) 
- Q\da_{x,+\mu} \gamma_5 Q_{x+\hat\mu,-\mu}
- Q\da_{x+\hat\mu,-\mu} \gamma_5 Q_{x,+\mu}].
\end{eqnarray}
The traces are over flavor and Dirac indices. We have introduced $8^2 = 64$ 
glueball operators
\begin{equation}
\Phi_{x,\pm\mu,\pm\nu} = \sum_i c^{i \dagger}_{x,\pm\mu} c^i_{x,\pm\nu},
\end{equation}
which satisfy the local commutation relations of $U(8)$
\begin{equation}
[\Phi_{x,\pm\mu,\pm\nu},\Phi_{y,\pm\rho,\pm\sigma}] = 
\delta_{xy} (\delta_{\pm\nu,\pm\rho} \Phi_{x,\pm\mu,\pm\sigma} 
- \delta_{\pm\mu,\pm\sigma} \Phi_{x,\pm\rho,\pm\nu}).
\end{equation}
We have also defined $(4 N_f)^2$ meson operators
\begin{equation}
M^{a \alpha b \beta}_x = \sum_i \Psi^{i a \alpha \dagger}_x \Psi^{i b \beta}_x,
\end{equation}
which carry flavor and Dirac indices and which generate the algebra of
$U(4 N_f)$
\begin{equation}
[M^{a \alpha b \beta}_x,M^{c \gamma d \delta}_y] = \delta_{xy}
(\delta_{bc} \delta_{\beta\gamma} M^{a \alpha d \delta}_x
- \delta_{ad} \delta_{\alpha\delta} M^{b \beta c \gamma}_x).
\end{equation}
While glueballs and mesons are not related via their commutation relations, 
i.e.
\begin{equation}
[\Phi_{x,\pm\mu,\pm\nu},M^{a \alpha b \beta}_y] = 0,
\end{equation}
they are both related with the constituent quark operators
\begin{equation}
Q^{a \alpha}_{x,\pm\mu} = \sum_i c^{i \dagger}_{x,\pm\mu} \Psi^{i a \alpha}_x,
\end{equation}
via the commutation relations
\begin{equation}
[\Phi_{x,\pm\mu,\pm\nu},Q^{a \alpha}_{y,\pm\rho}] = 
\delta_{xy} \delta_{\nu\rho} Q^{a \alpha}_{x,\pm\mu}, \
[M^{a \alpha b \beta}_x,Q^{c \gamma}_{y,\pm\mu}] = 
\delta_{xy} \delta_{ac} \delta_{\alpha\gamma} Q^{b \beta}_{x,\pm\mu}.
\end{equation}
Finally, the commutation relations of the constituent quark operators take the 
form
\begin{eqnarray}
&&[Q^{a \alpha}_{x,\pm\mu},Q^{b \beta \dagger}_{y,\pm\nu}] = 
\delta_{xy}(\delta_{ab} \delta_{\alpha\beta} \Phi_{x,\pm\mu,\pm\nu} 
- \delta_{\pm\mu,\pm\nu} M^{a \alpha b \beta}_x), \nonumber \\
&&[Q^{a \alpha}_{x,\pm\mu},Q^{b \beta}_{y,\pm\nu}] =
[Q^{a \alpha \dagger}_{x,\pm\mu},Q^{b \beta \dagger}_{y,\pm\nu}] = 0.
\end{eqnarray}
Thus, the inclusion of the constituent quark operators completes the site-based
algebra of $U(8) \otimes U(4N_f)$ to $U(8 + 4N_f)$. We expect this 
representation of the theory to be useful in the investigation of the large $N$
limit and in attempts to bosonize quantum link QCD.

\section{Conclusions}

We have constructed a new lattice formulation of QCD in the framework of quantum
link models. In these theories the link-based Hilbert space of the gluons is
finite. This is possible because the theory is constructed with a fifth
Euclidean direction that ultimately disappears via dimensional reduction,
although in the continuum limit its extent $\beta$ diverges in units of the
lattice spacing of the quantum link model. On the other hand, $\beta$ itself 
plays the role of the lattice spacing of another effective 4-d lattice theory, 
whose non-perturbatively generated length scales are exponentially large in 
$\beta$ (see fig.3).

An important question is whether $SU(N)$ quantum link models on a 4-d 
space-time lattice possess a continuum limit in the universality class of the
Coulomb phase of 5-d non-Abelian gauge theories. This is essential for their 
proper dimensional reduction to ordinary 4-d Yang-Mills theories. One way to
investigate this is to study the classical equations of motion of quantum link 
models. As will be shown in a forthcoming publication \cite{Bro97}, they indeed
resemble those of 5-d Yang-Mills theories with $A_5 = 0$. This shows that in 
the classical limit --- i.e. when one works with a large representation of 
$SU(2N)$ --- dimensional reduction does indeed take place. Whether dimensional 
reduction also occurs with the 20-dimensional representation of $SU(6)$ and 
hence whether the $SU(3)$ quantum link model introduced here provides a viable 
cut-off scheme for the continuum field theory of QCD, is an issue that can 
only be decided in numerical simulations. This situation is analogous to the 
one in quantum spin models. Based on the classical equations of motion one 
finds that for large spin the 2-d antiferromagnetic quantum Heisenberg model 
has an ordered ground state and that the corresponding Goldstone bosons --- in
that case antiferromagnetic magnons or spin-waves --- have a relativistic 
dispersion relation. Still, it required high-precision numerical simulations 
\cite{Wie94,Bea96} to be sure that a staggered magnetization is spontaneously 
generated even in the extreme quantum limit of spin $1/2$. Only this finally 
justifies the use of a 3-d relativistic low-energy chiral Lagrangian with 
$O(3)$ symmetry to describe the dynamics of the spin $1/2$ quantum Heisenberg 
model at low energies \cite{Has93}. Similarly, at present we cannot guarantee 
that the low-energy excitations of a 4-d $SU(3)$ invariant quantum link model 
formulated with the 20-dimensional representation of $SU(6)$ are 5-d massless 
gluons. However, this should at least be the case in the classical limit, i.e. 
when we work with a large representation of $SU(6)$. 

It is remarkable that the introduction of a fifth Euclidean direction of
finite extent also naturally protects the chiral symmetries of lattice
fermions, when open boundary conditions are used for the quarks at the two 4-d 
boundaries of the 5-d slab (see fig.2). This set-up --- first proposed by 
Shamir as a variant of Kaplan's fermion method --- fits very naturally with 
quantum link models. Due to tunneling between the two boundaries, a fermionic 
correlation length is generated, which is also exponentially large in $\beta$. 
Quantum link QCD treats quarks and gluons on an equal footing. They both have 
finite Hilbert spaces (per unit volume), both require a fifth Euclidean 
direction that finally disappears via dimensional reduction and they both have
correlation lengths that are exponential in $\beta$ without requiring 
fine-tuning. We believe that these properties point to an interesting 
connection between fermions and gauge fields in quantum link models. This may 
help in attempting to formulate chiral or supersymmetric gauge theories on the 
lattice.

Due to the fact that their gluonic Hilbert space is finite, quantum link models
can be expressed in terms of rishons, which are fermionic constituents of the 
gluons. The rishon dynamics can probably not be formulated as a continuum field
theory. Therefore we consider the rishons as lattice artifacts in our way to 
formulate QCD. The gluons in quantum link models are composites of rishons, 
just like antiferromagnetic magnons can be viewed as composites of electrons 
that are hopping on a lattice. Of course, in that case we know that the
electrons are real particles, which are hopping on a real crystal lattice. If 
our world is a thin five-dimensional slab with a lattice structure at extremely
small distances, the rishons in quantum link models could be real
particles. We do not want to speculate about that possibility and thus we
consider them as mathematical objects only. However, the rishons may turn out
to be of great importance for solving QCD in the large $N$ limit. We consider 
this possibility the most promising analytic aspect of the new formulation.

At finite $N$ we do not see a way to solve quantum link models analytically.
However, from a computational point of view quantum link QCD is still very
attractive, in particular because the new framework may allow one to construct 
very efficient cluster algorithms. The five-dimensional set-up is not as 
inconvenient as it may seem, because --- due to the discrete nature of the 
Hilbert space --- the 5-d path integral for quantum link models can be 
formulated and --- most important --- also be simulated directly in the 
continuum of the fifth Euclidean dimension. In fact, even in the standard 
formulation of lattice QCD it may turn out to be advantageous to work in five 
dimensions in order to control the chiral properties of the quarks.

It remains to be seen if quantum link models provide a more efficient 
formulation of QCD than standard lattice gauge theory. In any case, quantum 
link models allow us to attack the long-standing QCD problem from a different 
perspective.

\section*{Acknowledgements}

We are indebted to M. Basler, B. Beard, D. Chen, S. Levit, P. Orland, Y. Shamir
and A. Tsapalis for very interesting discussions. One of the  authors 
(U.-J. W.) likes to thank the theory group of the Weizmann Institute in 
Rehovot, where part of this work was done, for its hospitality and the A. P. 
Sloan foundation for its support.


\begin{thebibliography}{10}

\bibitem{Wil74}
K. Wilson, Phys. Rev. D10 (1974) 2445.

\bibitem{Nie96}
F. Niedermayer, Nucl. Phys. B (Proc. Suppl.) 53 (1997) 56.

\bibitem{Hor81}
D. Horn, Phys. Lett. 100B (1981) 149.

\bibitem{Orl90}
P. Orland and D. Rohrlich, Nucl. Phys. B338 (1990) 647.

\bibitem{Cha96}
S. Chandrasekharan and U.-J. Wiese, hep-lat/9609042, to appear in Nucl. Phys. B.

\bibitem{Kap92}
D. B. Kaplan, Phys. Lett. B288 (1992) 342.

\bibitem{Sha93}
Y. Shamir, Nucl. Phys. B406 (1993) 90.

\bibitem{Blu96}
T. Blum and A. Soni, hep-lat/9611030. 

\bibitem{Har79}
H. Harari, Phys. Lett. 86B (1979) 83.

\bibitem{Aue94}
A. Auerbach, ``Interacting Electrons and Quantum Magnetism'', Springer, 
New-York (1994).

\bibitem{Bea96}
B. B. Beard and U.-J. Wiese, Phys. Rev. Lett. 77 (1996) 5130.

\bibitem{Swe87}
R. Swendsen and S.-J. Wang, Phys. Rev. Lett. 58 (1987) 86.

\bibitem{Wol89}
U. Wolff, Phys. Rev. Lett. 62 (1989) 361; Nucl. Phys. B334 (1990) 581.

\bibitem{Has91}
P. Hasenfratz and F. Niedermayer, Phys. Lett. B268 (1991) 231.

\bibitem{Eve93}
H. G. Evertz, G. Lana and M. Marcu, Phys. Rev. Lett. 70 (1993) 875.

\bibitem{Wie94}
U.-J. Wiese and H.-P. Ying, Z. Phys. B93 (1994) 147.

\bibitem{Bea97a}
B. B. Beard, A. Ferrando, M. Greven and U.-J. Wiese, in preparation.

\bibitem{Bea97}
B. B. Beard, R. Brower, S. Chandrasekharan, A. Tsapalis and U.-J. Wiese, in 
preparation.

\bibitem{Wie93}
U.-J. Wiese, Phys. Lett. B311 (1993) 235.

\bibitem{Gal96}
A. Galli, hep-lat/9605026.

\bibitem{Fur95}
V. Furman and Y. Shamir, Nucl. Phys. 439 (1995) 54.

\bibitem{Has93}
P. Hasenfratz and F. Niedermayer, Z. Phys. B92 (1993) 91.

\bibitem{Lue83}
M. L\"uscher, Commun. Math. Phys. 85 (1982) 39.

\bibitem{Bro97}
R. Brower, S. Chandrasekharan and U.-J. Wiese, in preparation.

\end{thebibliography}
\end{document}